\documentclass[amsmath,amssymb,twocolumn,amsfonts,twocolumn]{revtex4}
\usepackage{graphicx}
\def \beq {\begin{equation}}
\def \eeq {\end{equation}}

\begin{document}
\title{Reactant-Product Quantum Coherence in Electron Transfer Reactions}
\author{I. K. Kominis}

\affiliation{Department of Physics, University of Crete, Heraklion 71103, Greece}

\begin{abstract}
We investigate the physical meaning of quantum superposition states between reactants and products in electron transfer reactions. We show that such superpositions are strongly suppressed and to leading orders of perturbation theory do not pertain in electron transfer reactions. This is because of the intermediate manifold of states separating the reactants from the products. We provide an intuitive description of these considerations with Feynman diagrams. We also discuss the relation of such quantum coherences to understanding the fundamental quantum dynamics of spin-selective radical-ion-pair reactions. \end{abstract}

\maketitle
\section{Introduction}
The role of quantum superpositions in irreversible chemical reactions, in particular electron transfer reactions \cite{et1}, has received little attention. The specific question we will here address is this: to first-order reaction kinetics,  can there be quantum superpositions between reactants and reaction products of an electron transfer reaction? 
\begin{figure}
\includegraphics[width=6.0 cm]{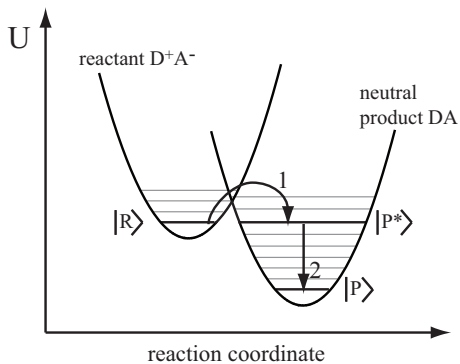}
\caption{(a) Level structure of reactants and reaction products in the electron transfer taking place in charge recombination of radical-ion pairs (reactants) and the formation of neutral products.}
\label{ET}
\end{figure}
Most if not all electron transfer reactions  are known to occur in two steps, as shown in Fig.\ref{ET}, depicting the potential curves of the reactants and the products: (1) the electron  tunnels from the initial state $|R\rangle$ into an intermediate state $|P^{*}\rangle$, which is a vibrational excited state of the product, the ground state of which is $|P\rangle$. (2) The excited state $|P^{*}\rangle$ decays to the ground state $|P\rangle$, with the emission of a photon or a phonon. This decay usually takes place at a rate much faster than the first step, so it is the latter that mostly determines the total reaction rate.  Most works on electron transfer deal with calculating the rate of this first step \cite{et2,et3,et4,et5,et6,et7}. Moreover, relevant considerations \cite{qcet} of quantum coherence have considered, again, this first step, and of course have appeared in the much broader context of chemical reactions \cite{brumer}. To our knowledge, however, the specific question we defined, i.e. whether the reactant state can be coherently coupled to the {\it final} product state in electron transfer reactions has not been addressed so far.

This question can be rephrased: Can states of the form $|\psi\rangle=c_{1}|{\rm R}\rangle+c_{2}|{\rm P}\rangle$ be physically realized {\it to first-order reaction kinetics}? The intuitive answer is negative, because of the existence of the intermediate states $|P^{*}\rangle$: a substantial amplitude for the product state $|P\rangle$ would translate to an equally suppressed amplitude of the intermediate $|P^{*}\rangle$ and thus a negligible amplitude of the reactant state $|R\rangle$. Here we are going to prove that the intuitive answer is indeed correct. In retrospect, this might seem to be an obvious statement.  

However, it can't be that obvious after all, for the following reasons. The seemingly innocuous question we will here address is directly related to understanding the fundamental quantum dynamics of spin-dependent recombination reactions of radical-ion pairs \cite{steiner}, which have recently received a lot of attention from the quantum information science perspective \cite{komPRE1,komPRE2,kom,JH,briegel,vedral,tiersch,sun,plenio}. In particular, the traditional theoretical description of spin-selective radical-ion-pair (RP) reactions has been questioned, and new approaches have been put forward by the authors in \cite{komPRE1,komPRE2, JH}, who introduced quantum measurements as conceptually important for understanding the quantum dynamics of these reactions. These papers have led to an ongoing debate \cite{JH,comment,reply,ivanov,purtov,tiersch} which at the least illuminates the richness of the topic in terms of the quantum mechanical concepts involved. Currently there exist at least three different theories, all purported to be derived from first principles, attempting to describe RP reactions. The traditional theory, sometimes called the Haberkorn master equation \cite{haberkorn}, was explicitly derived, for the first time to our knowledge, by Ivanov et al. \cite{ivanov}, and supported by Purtov along similar lines of thought \cite{purtov}. Both of these papers, however, start out from this seemingly innocent assumption: the existence of  quantum superpositions between the reactants (the radical-ion-pair state) and the reaction products (the neutral product state) of an electron transfer reaction. Furthermore, the considerations in \cite{vedral} also start out from the assumption that the radical-ion-pair state can be coherently coupled to the neutral product state. Thus, it is fair to say that what we are going to prove here is not an obvious, but a rather subtle physical statement regarding electron-transfer reactions in general, and radical-ion-pair reactions in particular.

In Section II we will rigorously prove our main assertion, namely that to {\it first order reaction kinetics}, the product state of an electron transfer reaction cannot be  coherently coupled to the reactant state, exactly due to the existence of intermediate states. In Section III we will elaborate on the physical consequences of neglecting these states, or stated equivalently, treating RP reactions by the Haberkorn master equation. We will show that according to the latter, the evolution of the entropy of the RP spin state exhibits an unphysical behaviour, not evidenced by the other two theories based on quantum measurement considerations. 
\section{Reactant-Product Quantum Coherence in Electron Transfer Reactions}
\begin{figure}
\includegraphics[width=6.0 cm]{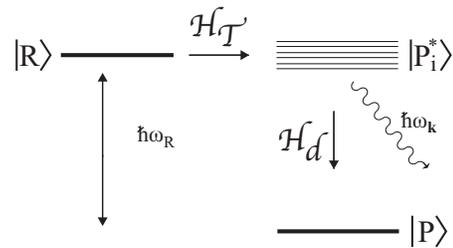}
\caption{Simplified level structure of an electron transfer process. The tunneling Hamiltonian ${\cal H}_{T}$ couples the "reactants" state $|R\rangle$ with the manifold of 
 excited states $|P^{*}\rangle$, which then radiatively decay to the ground state $|P\rangle$, the "product" state.}
\label{levels}
\end{figure}

We will consider a simplified version of the problem keeping all the ingredients that are essential for the basic physics. Reducing the problem to an electronic one, i.e. neglecting the nuclear degrees of freedom, we depict in Fig. \ref{levels}  the energy levels of the system comprised of the single reactant state $|R\rangle$, a manifold of intermediate states $|P^{*}_{i}\rangle$ to which the electron can tunnel from $|R\rangle$, and the single product state $|P\rangle$ to which $|P^{*}_{i}\rangle$ can decay by the emission of a photon or a phonon. Letting $|P\rangle$ define the zero of the energy scale, the unperturbed Hamiltonian of the system is 
\beq
{\cal H}_{0}=\omega_{R}a_{R}^{\dagger}a_{R}+\sum_{i}\omega_{P^{*}_{i}}a_{P^{*}_{i}}^{\dagger}a_{P^{*}_{i}}+\sum_{\mathbf{k}}\omega_{\mathbf{k}}a_{\mathbf{k}}^{\dagger}a_{\mathbf{k}}
\eeq
The sum over $\mathbf{k}$ denotes all modes of the photons (phonons) with frequency $\omega_{\mathbf{k}}=\omega_{R}$ due to energy conservation.
We will use fermionic operators to denote the creation ($a^{\dagger}$) or annihilation ($a$) of a single occupation of the relevant states.
The eigenstates of ${\cal H}_{0}$ are denoted by $|R,P^{*}_{i},P,N_{\mathbf{k}}\rangle$, where $R,P^{*}_{i},P=0,1$ are occupations of the reactant, the intermediate and the product states, respectively, and similarly $N_{\mathbf{k}}=0,1$ denotes the occupation of the photon state with wave vector $\mathbf{k}$. For example, the initial state is $|i\rangle=|1;0;0;0\rangle$, whereas the final state is one among the states  $|0;0;1;1_{\mathbf{k}}\rangle$, where $\mathbf{k}$ runs over all photon wave vectors satisfying energy conservation. These states will be collectively denoted by $|P\rangle$, i.e. without reference to the index $\mathbf{k}$. The tunneling Hamiltonian coupling $|R\rangle$ to $|P^{*}_{i}\rangle$ is simply
${\cal H}_{T}=\sum_{i}\lambda_{i} a_{R}a_{P^{*}_{i}}^{\dagger}\label{tun}+c.c.$
Finally, the decay Hamiltonian coupling $|P^{*}_{i}\rangle$ to the radiation (or phonon) reservoir and leading to the ground state $|P\rangle$ is 
${\cal H}_{d}=\sum_{i}\sum_{\mathbf{k}} c_{i,\mathbf{k}}a_{P^{*}_{i}}a_{P}^{\dagger}a_{\mathbf{k}}^{\dagger}+c.c.\label{rad}$
The coupling constants $c_{i,\mathbf{k}}$ readily result from the quantum theory of light-matter interaction, and are $c_{i,\mathbf{k}}=\sqrt{\hbar\omega_{\mathbf{k}}/2V}{\cal M}_{i}$, where $V$ is the volume in which the radiation field modes are defined, and ${\cal M}_{i}$ the relevant matrix element for the transition $P^{*}_{i}\rightarrow P$. The total interaction Hamiltonian is ${\cal V}_{int}={\cal H}_{T}+{\cal H}_{d}$. It is worth considering two simple cases before proceeding:  (i) Neglecting the presence of $|P\rangle$ and considering only the tunneling transition $|R\rangle\rightarrow |P^{*}_{i}\rangle$, moreover assuming that $\lambda_{i}=\lambda$ for all $i$, and using Fermi's golden rule, it is found within $1^{\rm st}$-order perturbation theory that this transition's rate is $k=(2\pi/\hbar^{2})|\lambda|^{2}\rho^{*}(\omega_{R})$, where $\rho^{*}(\omega)$ is the density of states of the $\{P^{*}\}$-manifold calculated at the energy of the reactants $\omega_{R}$.  (ii) Neglecting the presence of $|R\rangle$ and considering only the radiative transition $|P^{*}_{i}\rangle\rightarrow |P\rangle$, moreover assuming that ${\cal M}_{i}={\cal M}$ for all $i$, and again using Fermi's golden rule, it is found within $1^{\rm st}$-order perturbation theory that this transition's rate is $\Gamma=(2\pi/\hbar^{2})|{\cal M}|^{2}\rho(\omega_{P^{*}})$, where $\rho(\omega)$ is the density of states of the radiation field calculated at the energy of the excited intermediates $\omega_{P^{*}}$. It is noted that in realistic systems it usually is $\Gamma\gg k$. 

We will now move to consider the action of ${\cal V}_{int}$ on the entire system using time-dependent perturbation theory. The initial state $|R\rangle=|1;0;0;0\rangle$ will evolve to $|\psi_{t}\rangle=c_{R}(t)|R\rangle+\sum_{i}c_{P^{*}_{i}}(t)|P^{*}_{i}\rangle+c_{P}(t)|P\rangle$. To $0^{\rm th}$-order perturbation theory it is $c_{R}^{(0)}(t)=1$ and $c_{P^{*}_{i}}^{(0)}(t)=c_{P}^{(0)}(t)=0$.
It is clear that within $1^{\rm st}$-order perturbation theory,  ${\cal V}_{int}$ can couple $|R\rangle$ to $|P^{*}_{i}\rangle$ through ${\cal H}_{T}$, but not to $|P\rangle$, hence $c_{P}^{(1)}=0$. Let's now move to $2^{\rm nd}$-order perturbation theory, where a possible coupling between $|R\rangle$ and $|P\rangle$ appears through the product ${\cal H}_{d}{\cal H}_{T}$.  Assuming that the reactant and intermediate states are non-degenerate, i.e. $\omega_{R}\neq \omega_{P^{*}_{i}}$ it follows from $2^{\rm nd}$-order time-dependent perturbation theory (setting $\delta_{\mathbf{k}}=\omega_{\mathbf{k}}-\omega_{R}$) that
\begin{align}
c_{P}^{(2)}&={{it}\over \hbar^{2}}\sum_{\mathbf{k}}e^{i\delta_{\mathbf{k}}t/2}{\rm sinc}\Big[\delta_{\mathbf{k}}t/2\Big]\sum_{i}{{{\cal V}_{Pi}{\cal V}_{iR}}\over {\omega_{P^{*}_{i}}-\omega_{R}}}\nonumber\\
&={{it}\over \hbar^{2}}\sum_{\mathbf{k}}e^{i\delta_{\mathbf{k}}t/2}{\rm sinc}\Big[\delta_{\mathbf{k}}t/2\Big]\sum_{i}{{\lambda_{i}c_{i,\mathbf{k}}}\over {\omega_{P^{*}_{i}}-\omega_{R}}}\label{c2}
\end{align}
As the manifold of intermediate states is a dense quasi-continuum, and as mentioned before, $\lambda_{i}=\lambda$ and $c_{i,\mathbf{k}}=c_{\mathbf{k}}$ for all $i$ can be taken out of the sum, moreover since $1/(\omega_{P^{*}_{i}}-\omega_{R})$ is an odd function of $\omega_{P^{*}_{i}}-\omega_{R}$, it follows that $c_{P}^{(2)}=0$.
Even in the case of resonant tunneling, i.e. supposing the existence of one intermediate state $|P^{*}\rangle$ resonant with $|R\rangle$, it follows that 
\begin{align}
c_{P}^{(2)}&=-{1\over \hbar^{2}}\lambda\sum_{\mathbf{k}}c_{\mathbf{k}}{{e^{i\delta_{\mathbf{k}}t}(1-i\delta_{\mathbf{k}}t)-1}\over {\delta_{\mathbf{k}}^{2}}}\nonumber\\
&=-{t^{2}\over \hbar^{2}}\lambda\sum_{\mathbf{k}}c_{\mathbf{k}}e^{i\delta_{\mathbf{k}}t}
\end{align}
where the second equality follows in the limit $\delta_{\mathbf{k}}\rightarrow 0$ imposed by energy conservation. The phase $e^{i\delta_{\mathbf{k}}t}$ will average $ c_{P}^{(2)}$ to zero. Even if $\delta_{\mathbf{k}}=0$ identically, this contribution is not first-order in the reaction kinetics, but third, so it can be neglected.

\begin{figure}
\includegraphics[width=6.0 cm]{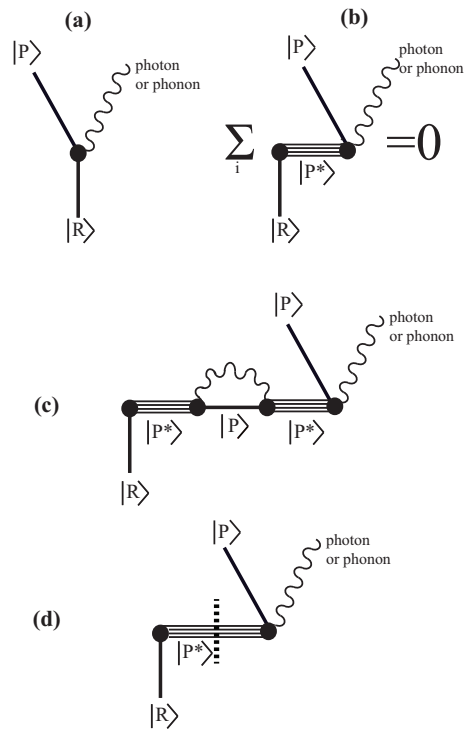}
\caption{(a) "Effective" description of electron-transfer reactions neglecting the intermediate states and assuming a direct coupling between reactants and products. (b) Taking into account the intermediate states, this is the lowest-order ($2^{\rm nd}$) coherent coupling between reactants and products, the amplitude of which is zero. (c) The first non-zero amplitude comes from this $4^{\rm th}$-order diagram. (d) Within frist-order reaction kinetics, the reaction proceeds by a real transition from $|R\rangle$ to $|P^{*}\rangle$ (indicated by the dashed line), followed by another real transition from $|P^{*}\rangle$ to $|P\rangle$.}
\label{feyn}
\end{figure}

Now it is also clear that coherently coupling $|P\rangle$ to $|R\rangle$ is possible only within even-order perturbation theory, hence the first non-zero term in the sought after quantum superposition appears at $4^{\rm th}$-order perturbation theory, easily found to be negligible.
There is a very transparent way to summarize the above results with Feynman diagrams, shown in Fig.\ref{feyn}. Not taking into account the intermediate states $|P^{*}\rangle$, i.e. supposing a physically absent direct coherent coupling between $|R\rangle$ and $|P\rangle$, is depicted by the diagram shown in Fig.\ref{feyn}a. Taking the full level dynamics into account, we have shown that the $2^{\rm nd}$-order perturbation theory diagram depicted in Fig.\ref{feyn}b has a zero amplitude due to the energy-odd denominator.  The first non-zero amplitude for coherently coupling $|R\rangle$ to $|P\rangle$ results from the $4^{\rm th}$-order diagram of Fig.\ref{feyn}c. In this diagram, the intermediate state $|P^{*}\rangle$ emits a virtual photon that is re-absorbed by the produced product state, leading again to the intermediate manifold before the actual product is generated. Finally, the electron transfer reaction actually proceeds within first order reaction kinetics through two consecutive {\it real} transitions, i.e. there is a {\it real} transition from $|R\rangle$ to $|P^{*}\rangle$, followed by another {\it real} transition from $|P^{*}\rangle$ to $|P\rangle$. We symbolize this with the diagram shown in Fig.\ref{feyn}d, where the dashed line crossing the $|P^{*}\rangle$ states is supposed to mean that the $|P^{*}\rangle$ states are really, as opposed to virtually, populated by the real transitions from $|R\rangle$.  
\section{Neglecting intermediate states}
\begin{figure}
\includegraphics[width=6.0 cm]{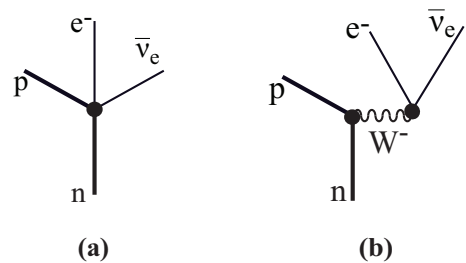}
\caption{(a) Early understanding of neutron decay, leading to several problems, such as an infinite electron-neutrino scattering cross section. (b) The infinity was remedied when the actual vertex involving the intermediate W$^{-}$ boson was understood from electroweak theory.}
\label{weak}
\end{figure}

As mentioned in the introduction, the authors in \cite{ivanov} and \cite{purtov} derived the traditional master equation of spin chemistry treating the recombination process according to the diagram in Fig. \ref{feyn}a, i.e. by neglecting the intermediate states $|P^{*}\rangle$. This is an example of what field theory calls an "effective theory". There are many examples in physics, especially in nuclear physics \cite{nuc}, where effective theories are indeed a good approximation, destined however, to fail in some regime of the relevant parameter space. For example, weak interactions were initially described by the effective Fermi theory, according to which the neutron decays as shown in Fig.\ref{weak}a. 
After the development of electroweak theory, the relevant Feynman diagram was found to include the intermediate W$^{-}$ boson (Fig.\ref{weak}b). The Fermi theory was a good low energy approximation, plagued, however, by several problems, such as an electron-neutrino scattering cross section tending to infinity at high energies. This and other problems were remedied by the full electroweak theory.
Similarly, the traditional master equation of spin chemistry is a very good approximation, plagued however, with several conceptual problems.  We will now turn to those. 

We consider a very simple example of a radical-pair consisting of just two unpaired electrons and no nuclear spins. We furthermore set to zero all magnetic interactions and consider a single recombination channel, e.g. the singlet, with $k_S$ being the recombination rate. We take as an initial state the maximally coherent state $|\psi_{0}\rangle=(|S\rangle+|T_{0}\rangle)/\sqrt{2}$, where $|S\rangle$ and  $|T_{0}\rangle$ are the two-electron singlet and zero-projection triplet state, respectively. 
\begin{figure}
\includegraphics[width=7.5 cm]{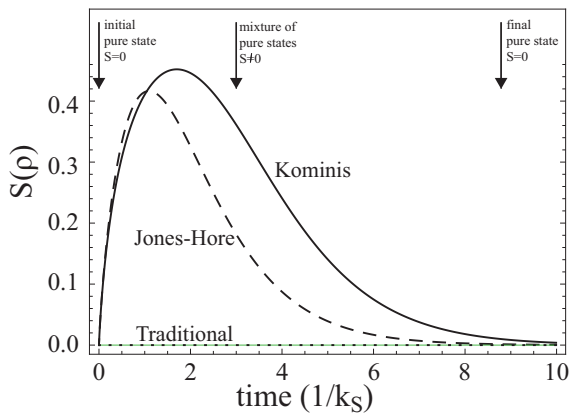}
\caption{Comparison of the three master equations by Kominis (solid line), Jones-Hore (dashed line) and the traditional (dotted line) in predicting the von-Neuman entropy $S(\rho)$. The density matrix $\rho$ was normalized by ${\rm Tr}\{\rho\}$ in order to remove the artefact of diminishing RP population due to recombination.}
\label{comparison}
\end{figure}
What is physically obvious is that a fraction of the radical-ion pairs, initially in the state $|\psi_{0}\rangle$ will recombine through the singlet channel, while the rest will remain locked in the non-reactive triplet state forever (we took $k_{T}=0$). Indeed this is what the prediction of all three theories is, albeit with a quantitative difference of what exactly this fraction is \cite{komPRE2}. This difference is immaterial for the current consideration. What is also expected on physical grounds is the fact that while at the beginning and the end the state of the radical-ion pairs is pure, i.e. at $t=0$ we have a pure singlet state and at $t\rightarrow\infty$ we have a pure triplet, during intermediate times we have a mixture of pure states. Hence one would expect that the von-Neuman entropy $S(\rho)=-{\rm Tr}\{\rho\ln\rho\}$, where $\rho$ is the density matrix describing the two-electron spin state, would start from zero, acquire a non-zero value at intermediate times, and return to zero. In Fig.\ref{comparison}a we compare the predictions of the three master equations. It is clear that the two theories based on the quantum measurement approach to the recombination of RPs predict the above described behavior expected on general physical grounds. In contrast, $S(\rho)=0$ at all times according to the traditional theory.

The aforementioned problem of the Haberkorn master equation can be traced to the absence of intermediate states in the theoretical paradigm underlying the traditional theory. The presence of such states leads to a fundamental and unavoidable S-T dephasing within first order reaction kinetics through Feynman diagrams like the ones presented in \cite{komPRE2}. Concluding, we have explored the physical existence of quantum superpositions between reactants and products in electron transfer reactions and connected the result to the fundamental quantum dynamics of spin chemistry reactions.

\end{document}